\documentclass[a4paper]{jpconf}
\usepackage{graphicx}
\begin{document}
\title{Pitfalls in the analysis of low-temperature thermal conductivity of 
high-$T_c$ cuprates}

\author{Yoichi Ando$^1$, X. F. Sun$^2$, and Kouji Segawa$^1$}

\address{$^1$ Institute of Scientific and Industrial Research,
Osaka University, Ibaraki, Osaka 567-0047, Japan}

\address{$^2$ Hefei National Laboratory for Physical Sciences at
the Microscale, University of Science and Technology of China,
Hefei, Anhui 230026, P. R. China}

\begin{abstract}

Recently, it was proposed that phonons are specularly reflected below
$\sim$0.5 K in ordinary single-crystal samples of high-$T_c$ cuprates,
and that the low-temperature thermal conductivity $\kappa$ should be
analyzed by fitting the data up to $\sim$0.5 K with the formula $\kappa
= aT + bT^{\alpha}$ where $\alpha$ is a fitting parameter. Such an
analysis yields a result different from that obtained from the
conventional analysis in which the fitting is restricted to a region
below $\sim$0.15 K, where $\kappa$ has been found to obey $\kappa = aT +
bT^3$ very well, which means that phonons are in the boundary scattering
regime without specular reflection. Here we show that the proposed new
analysis is most likely flawed, because the specular phonon
reflection means that the phonon mean free path $\ell$ gets {\it longer}
than the mean sample width, while the estimated $\ell$ is actually much
{\it shorter} than the mean sample width above $\sim$0.15 K. 

\end{abstract}

\section{Introduction}

Because high-$T_c$ cuprate superconductors have a $d$-wave
superconducting gap that has four ``nodes" where the gap vanishes, there
are low-energy quasiparticles (QPs) in the superconducting state of
cuprates \cite{HusseyRev}. Those QPs, called nodal QPs, are extended and
have well-defined wave vectors, and they can carry heat even at very low
temperature. The thermal conductivity $\kappa$ of cuprates below $T_c$
has turned out to be a useful probe for studying the nature of the
superconducting state, because the physics of the nodal QPs are directly
reflected in $\kappa$. In particular, at very low temperature where the
QPs are in the residual elastic scattering regime, the electronic
thermal conductivity $\kappa_e$ is expected to be approximated by a
``universal" thermal conductivity $\kappa_0$, which is expressed as
\cite{Graf,Durst}
\begin{equation}
\frac{\kappa_0}{T} = \frac{k_B^2}{3 \hbar} \frac{n}{c} \left(
\frac{v_F}{v_2} + \frac{v_2}{v_F} \right) \simeq \frac{k_B^2}{3
\hbar} \frac{n}{c} \frac{v_F}{v_2}, \label{k0/T}
\end{equation}
where $n$ is the number of $\rm{CuO_2}$ planes per unit cell, $c$ is the
$c$-axis lattice constant, and $v_F$ ($v_2$) is the QP velocity normal
(tangential) to the Fermi surface at the gap node. This $\kappa_0$ is
called ``universal" in the sense that it is expected to be independent
of the QP scattering rate $\gamma$; this happens because the effects of
disorder to create QPs and to scatter them would exactly cancel each
other in a certain range of $\gamma$ \cite{HusseyRev}. Thus, the bulk
measurement of $\kappa_e$ at low enough temperature would give us
microscopic information about the gap. 

However, in metals both electrons and phonons carry heat, so $\kappa$ is
a sum of $\kappa_e$ and the phonon thermal conductivity $\kappa_p$. This
causes a fundamental complication in the data analysis, that is, one has
to separate $\kappa_e$ from $\kappa_p$ in the measured $\kappa$. Such a
separation can be accomplished if $\kappa_e$ and $\kappa_p$ have
distinct and well-defined temperature dependences; actually, at low
enough temperature, $\kappa_e \simeq aT$ is expected in the residual
elastic scattering regime, and $\kappa_p \simeq bT^3$ is expected in the
boundary scattering regime of phonons. Hence, experimentalists often
employ dilution refrigerators to measure the thermal conductivity in the
mK region, where the heat transport is usually found to be in the
above-described regime. Indeed, in the case of cuprates, we have
recently shown \cite{logT} that the description $\kappa = aT + bT^3$
holds very well in superconducting YBa$_2$Cu$_3$O$_y$ (YBCO) samples
below $\sim$110 mK.

Recently, there is a controversy in the high-$T_c$ community as to how
best the $\kappa(T)$ data at low temperature are analyzed to separate
$\kappa_e$ from $\kappa_p$. In a paper reporting the low-temperature
$\kappa(T)$ behavior in YBCO and La$_{2-x}$Sr$_x$CuO$_4$ (LSCO),
Sutherland {\it et al.} argued \cite{Sutherland_PRB} that in
single-crystal samples of those materials, the specular reflection of
phonons becomes important at temperatures below $\sim$0.5 K, and hence
the analysis of the low-temperature data should take this effect into
account. The new analysis proposed in Ref. \cite{Sutherland_PRB}
was adopted in subsequent papers from the same group, such as Refs.
\cite{Hill, Li, Sutherland_PRL, newPRL, newLi}, and has been
playing crucial roles in deducing the conclusions of those papers. In
this article, we point out that the key assumption behind the new analysis,
that the phonons are specularly reflected at $T$ up to $\sim$0.5 K, is
erroneous, by showing that the phonon mean free path is much {\it
shorter} than the mean sample width in most of the proposed temperature
range, which means that phonons are not even in the boundary
scattering regime that is a prerequisite for the dominance of specular
reflections.

\section{Specular Phonon Reflection?}

The standard way of analyzing the low-$T$ thermal conductivity data
$\kappa(T)$ in terms of $\kappa/T = a + b T^2$, which was first employed
in the present context by Taillefer {\it et al.} \cite{Taillefer} for
both insulating and superconducting YBCO, has proved to be reliable for
all cuprate systems studied so far \cite{Hussey, Takeya, Ando, Sun1,
Sun2, Behnia}. The physical ground of this analysis is that the $T$
dependence of $\kappa$ does not change further at lower $T$ when the
electrons are elastically scattered and the phonons are scattered by the
boundary, and hence one can safely assume the formula $\kappa/T = a + b
T^2$ to be valid down to $T$ = 0 K, which gives confidence in the $T
\rightarrow 0$ extrapolation; note that in the boundary scattering
regime the phonons are usually diffusively scattered at the sample
surface \cite{Hurst} and the phononic contribution $\kappa_p$ is simply
proportional to the specific heat, yielding $\kappa_p = bT^3$. However,
Sutherland {\it et al.} had advocated in their paper
\cite{Sutherland_PRB} that the specular reflection of phonons causes a
$T$ dependence of $\kappa_p$ weaker than $T^3$ in the boundary
scattering regime, and that a fitting of the data for a wider
temperature range to $\kappa/T = a + b T^{\alpha-1}$ with $\alpha <$ 3
is better than the standard one. In the following, since the analysis of
the $\kappa(T)$ data of underdoped YBCO has been the most controversial
\cite{logT, Sutherland_PRL, newPRL} after the publication of Ref.
\cite{Sutherland_PRB}, we focus our discussion on the case of YBCO.

\begin{figure}
\begin{center}
\includegraphics[clip,width=9cm]{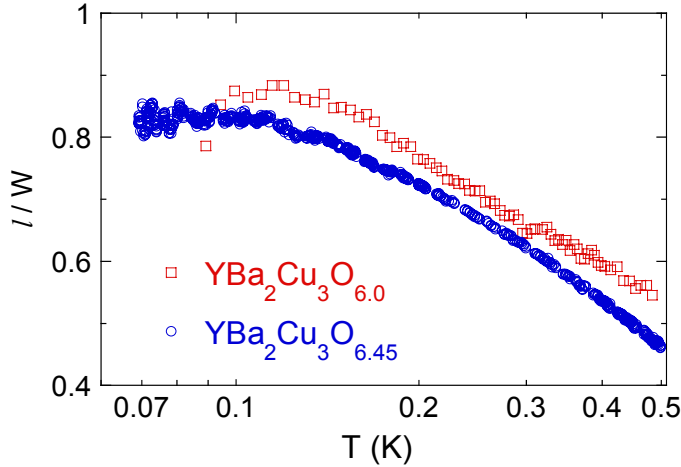}
\end{center}
\caption{Temperature dependence of the calculated
phonon mean free path $\ell$ divided by the mean sample width
$W$ for YBCO. The data for $y$ = 6.0 and 6.45 are taken from Refs.
\cite{Taillefer} and \cite{logT}, respectively.}
\end{figure}

When the phonons are specularly reflected at the surface, $\kappa_p$
becomes dependent on the averaged phonon wave length that changes with
$T$, and empirically $\kappa_p$ is expressed as $bT^{\alpha}$ with
$\alpha <$ 3 \cite{Hurst,Pohl}. Past studies on insulating materials
unanimously found \cite{Hurst,Pohl,Thacher} that the phonon mean free
path $\ell$ becomes much {\it longer} than the mean width $W$ of the
sample when the specular reflection becomes important, as is naturally
expected ($W$ is usually taken to be $2/\sqrt{\pi}$ times the
geometrical mean width $\bar{w}$, see Ref. \cite{Taillefer}). Therefore,
calculating $\ell$ in YBCO and comparing it with $W$ is a good way of
judging whether the phonons are indeed specularly reflected. This can be
rather easily done, because $\kappa_p$ is equal to
$\frac{1}{3}C\bar{v}\ell$ where $C = \beta T^3$ is the phonon specific
heat and $\bar{v}$ is the averaged sound velocity, and both $\beta$ and
$\bar{v}$ are known for YBCO \cite{Taillefer}. Figure 1 shows the
temperature dependence of $\ell$ calculated with the data reported by
Taillefer {\it et al.} \cite{Taillefer} for $y$ = 0 (where $W$ was
specified) and with our own data \cite{logT} for $y$ = 6.45 \cite{note}.
One can see that $\ell$ becomes saturating and comparable to $W$ only at
the lowest temperature, and it stays {\it smaller} than $W$ in the
temperature range where Sutherland {\it et al.} argued
\cite{Sutherland_PRL, Sutherland_PRB} that the specular reflection is
taking place (0.07 -- 0.5 K). Therefore, Fig. 1 demonstrates that the
assertion of Sutherland {\it et al.} is dubious and the specular
reflection is not important in YBCO. A natural conclusion is that in
YBCO at temperatures higher than $\sim$120 mK phonons are {\it not} in
the boundary scattering regime \cite{Taillefer, Sun1, Sun2} and the
phenomenological $T^{\alpha}$ ($\alpha <$ 3) dependence comes from some
additional scattering (such as scatterings off point defects,
dislocations, twin boundaries, etc.) that only reduces $\ell$
\cite{Berman}; such additional scattering would eventually die away at
lower $T$ as the boundary scattering regime is achieved, so the
$T^{\alpha}$ fitting should not be used for the $T \rightarrow$ 0
extrapolation. 

\section{Validity of Extrapolations}

As a matter of fact, one must always bear in mind that, if an
extrapolation is to be employed in the analysis, {\it the formula used
for the extrapolation should have a physical reasoning which assures
that the same functional form holds throughout the range of the
extrapolation}. Otherwise, it is always possible that the true
temperature dependence of the system changes below the experimental
range (due, for example, to a change in the dominant scattering
mechanism) and the extrapolation gives an inadequate estimate for the
$T$ = 0 value. In the present case, when the assumption of the specular
phonon reflection fails, there is no physical ground to expect the
phenomenological $T^{\alpha}$ dependence of $\kappa_p$ to continue to
$T$ = 0, and hence the conclusion about the electronic contribution at
$T$ = 0 loses its footing.

\section{Conclusion}

We have shown that in high-quality YBCO single crystals, the phonons are
{\it not} likely to be specularly reflected at $\sim$0.5 K because the
estimated phonon mean free path is much {\it shorter} than the mean
sample width at such relatively high temperature. Hence, the data
analysis assuming the specular phonon reflection lacks its physical
ground. When analyzing the low-temperature thermal conductivity data,
one should avoid falling into pitfalls of fitting the data using some
accidental power law and extrapolating it to zero temperature, when
there is no particular reason to expect to its validity.

\section*{Acknowledgment}

Y.A. was supported by KAKENHI 16340112 and 19674002 provided by the
Japan Society for the Promotion of Science, and X.F.S. acknowledges the
support from the National Natural Science Foundation of China (10774137
and 50421201) and the National Basic Research Program of China
(2006CB922005).

\end{document}